\documentclass[reprint,amsmath,amssymb,aps,prb]{revtex4-1}
\usepackage[T1]{fontenc}
\usepackage{array,booktabs}
\usepackage{graphicx}
\usepackage{mathrsfs}
\usepackage{bm}
\usepackage{hyperref}
\usepackage{dcolumn}
\usepackage{amsmath}
\usepackage{amssymb}
\usepackage{xcolor}
\usepackage{subfig}
\usepackage{caption}
\usepackage{siunitx}
\usepackage{ulem}
\usepackage{textcomp}
\usepackage[version=4]{mhchem}
\hypersetup{colorlinks=true, linkcolor=blue, citecolor=red, urlcolor=blue}

\everymath{\displaystyle}

\newcommand{\kk}{\bm{k}}

\newcommand{\pp}{\bm{p}}
\newcommand{\qq}{\bm{q}}
\newcommand{\rr}{\bm{r}}
\newcommand{\xx}{\bm{x}}
\newcommand{\GG}{\bm{G}}
\newcommand{\KK}{\bm{K}}
\newcommand{\LL}{\bm{L}}
\renewcommand{\SS}{\bm{S}}
\newcommand{\ttau}{\bm{\tau}}

\newcommand{\Abs}[1]{\left|{#1}\right|}
\newcommand{\Mean}[1]{\left\langle{#1}\right\rangle}

\newcommand{\up}{\uparrow}
\newcommand{\down}{\downarrow}

\newcommand{\bfQ}{\mathbf{Q}}
\newcommand{\bfT}{\mathbf{T}}
\newcommand{\mcT}{\mathcal{T}}
\newcommand{\mcI}{\mathcal{I}}

\newcommand{\Braoperket}[3]{\left\langle{#1}\middle|{#2}\middle|{#3}\right\rangle}

\begin{document}

\title{Magnetic States of Graphene Proximitized Kitaev Materials}
\author{Jingtian Shi and A.H. MacDonald}
\affiliation{Department of Physics, University of Texas at Austin, Austin TX 78712, USA}
\date{\today}

\begin{abstract}

Single layer \ce{\alpha}-ruthenium trichloride (\ce{\alpha-RuCl_3}) has been proposed as a potential quantum spin liquid.
Graphene/\ce{RuCl_3} heterobilayers have been extensively studied with a focus on the large interlayer 
electron transfer that dopes both materials. 
Here we examine the interplay between the competing magnetic state of \ce{RuCl_3} layer and graphene electronic properties.
We perform self-consistent Hartree-Fock calculations on a Hubbard-Kanamori model of the $4d^5$ $t_{2g}$ 
electrons of \ce{\alpha-RuCl_3} and confirm that out-of-plane ferromagnetic and zigzag antiferromagnetic states are energetically competitive. We show that the influence of hybridization between graphene and \ce{RuCl_3} bands is strongly sensitive to
the magnetic configuration of \ce{RuCl_3} and the relative orientations of the two layers.  We argue that 
strong hybridization leads to graphene magneto-resistance and that it may tilt the balance between closely competing 
magnetic states.  Our analysis can be applied to any van der Waals heterobilayer system with weak interlayer 
hybridization and allows for arbitrary lattice constant mismatch and relative orientation.  
\end{abstract}

\maketitle

\section{Introduction}

The discovery of two-dimensional (2D) intrinsic ferromagnets \cite{huang2017layerdependent, gong2017discovery} opened up van der Waals magnetism \cite{gibertini2019magnetic, li2019intrinsic} as a promising research topic. 
Given a bulk van der Waals magnet platform, a wide range of magnetic and spintronic properties can 
often be flexibly engineered in its atomically thin films.  The case of van der Waals 
Kitaev materials \cite{trebst2022kitaev} like \ce{\alpha}-ruthenium trichloride (\ce{\alpha-RuCl_3}) \cite{banerjee2016proximate}
is particularly intriguing because of 
their potential to realize highly unusual quantum spin liquid phases \cite{savary2017quantum, zhou2017quantum, hermanns2018physics, knolle2019field}. 
The term Kitaev material refers to Mott insulators that are described by frustrated spin Hamiltonians
with large Kitaev components \cite{kitaev2006anyons}, whether or not they actually realize the spin-liquid state. 
The magnetic ion sublattices of bulk Kitaev materials usually have honeycomb lattice layers populated by 
heavy $d^{5}$ transition metal ions \cite{jackeli2009mott}. 

Thanks to the development of the tear-and-stack techniques \cite{kim2016van, cao2016superlatticeinduced} for van der Waals compounds, single-layer and few-layer \ce{RuCl_3} structures can be  
flexibly stacked with other van der Waals compounds to form a new class of tunable 
2D multilayers \cite{biswas2019electronic, gerber2020initio, mashhadi2019spinsplit, zhou2019evidence, rizzo2020chargetransfer, wang2020modulation, wang2022direct, rizzo2022nanometerscale, balgley2022ultrasharp, yang2023magnetic, zheng2023tunneling}.  The work function difference between 
graphene and \ce{RuCl_3} results in a large electron density transfer from graphene to \ce{RuCl_3}.
Measured graphene hole densities vary across experiments, ranging
between $0.03\sim 0.06e$ per Ru atom ($2\sim 4 \times 10^{13} \rm cm^{-2}$)\cite{rizzo2020chargetransfer, wang2020modulation}. 
(The sample dependence is not yet fully understood.)
Electrical gates can induce more limited changes in layer resolved charge density, 
providing a route to control the doping of 2D Mott insulators without introducing disorder.  

Bulk \ce{\alpha-RuCl_3} has zigzag antiferromagnetic (AFM) order at low temperature, as confirmed by both neutron scattering experiments \cite{sears2015magnetic, johnson2015monoclinic, cao2016lowtemperature, banerjee2017neutron} and \textit{ab initio} simulations \cite{kim2015kitaev, winter2016challenges, wang2017theoretical}. The magnetic state is believed to result from competition between the Kitaev interaction and various non-Kitaev terms in the effective spin model \cite{rau2014generic, winter2016challenges}.
For monolayer \ce{\alpha-RuCl_3}, however, numerical simulations can predict
ferromagnetic (FM) \cite{sarikurt2018electronic} or zigzag AFM \cite{gerber2020initio} states, depending on details.
All \textit{ab initio} electronic structure calculations 
agree that the two states are very close in energy. The effect of an adjacent graphene layer has variously been reported to 
favor zigzag states \cite{gerber2020initio, souza2022magnetic} or to enhance the Kitaev term in the spin model and suppress
non-Kitaev terms -- moving the system towards \cite{biswas2019electronic, gerber2020initio} the spin-liquid region of its phase diagram. 
However, the large lattice constant mismatch between graphene and \ce{\alpha-RuCl_3} complicates electronic structure simulations.
At the same time, experimental probes of the single-layer magnetic state are challenged by the 
inapplicability of neutron scattering. One possible method to probe 2D magnetism directly
is to employ a superconducting quantum interference device (SQUID), which is a powerful tool able to detect out-of-plane ferromagnetism in 2D materials, however, only at a temperature lower than $1 \sim 2 \rm K$. Spin-resolved photoemission measurements are also difficult due to the limited size of exfoliated flakes.  One promising alternative is to use optical probes \cite{zhong2017van}.  Still, the true magnetic ground 
state of the \ce{RuCl_3} layer remains a mystery at present.

In this work we propose that the transport properties of \ce{RuCl_3}-proximitized graphene can be used as an indirect probe of 
the \ce{RuCl_3} magnetic state.
In Sec. \ref{sec_model} we introduce the microscopic Hamiltonian we use to model the monolayer \ce{RuCl_3}.
In Sec. \ref{sec_results} we present the results of a Hartree-Fock simulation on the Mott insulating state of monolayer \ce{RuCl_3}, including the magnetic configurations and the electronic conduction band structures of various metastable magnetic states. We have reproduced both FM and zigzag AFM states as well as their close competition in energy. We also find a new metastable state with $\sqrt{3} \times \sqrt{3}$ magnetic cells, which is fragile to small variations of model parameters.
In Sec. \ref{sec_int_with_graphene} we analyze the effect of coupling between \ce{RuCl_3} and graphene bands in the heterobilayer allowing for arbitrary relative orientation between the layers, recognizing that interlayer hybridization is weak compared to other energy scales, and that intersections of the isolated layer Fermi surfaces play a key role.
We find that in both zigzag AFM and $\sqrt{3} \times \sqrt{3}$ states, within certain ranges of twist angle, the Fermi surfaces of the two materials intersect, allowing for an interlayer-coupling-induced avoided-crossing gap
to open near the Fermi energy, and argue that this can cause a substantial increase in resistivity of the system.
On the contrary, in FM state the Fermi surfaces of the two layers never intersect.
Our treatment of the interlayer hybridization is inspired by the Bistritzer-MacDonald model of twisted bilayer graphene \cite{bistritzer2011moire} and can be applied to other 2D materials with large mismatches in lattice constant and orientation.
We develop a phenomenological model in which the interlayer coupling contribution to the Hamiltonian is governed by three free parameters by applying symmetry restrictions.  We calculate the band structure of the bilayer system 
and indeed see the avoided band crossing near the Fermi energy. The bands of isolated graphene and zigzag-state \ce{RuCl_3} are doubly degenerate as a result of a Kramer's degeneracy protected by combined 3-dimensional (3D) spatial inversion and time-reversal symmetry ($\mcI\mcT$). The degeneracy is lifted at the avoided crossings by $\mcI\mcT$-violating interlayer coupling, leading to nonzero net magnetic moment and spin-dependent transport properties.
In Sec. \ref{sec_summary} we discuss and propose possible future developments.

\section{Model}
\label{sec_model}

\begin{figure}[tb]
	\centering
	\includegraphics[width=0.5\textwidth]{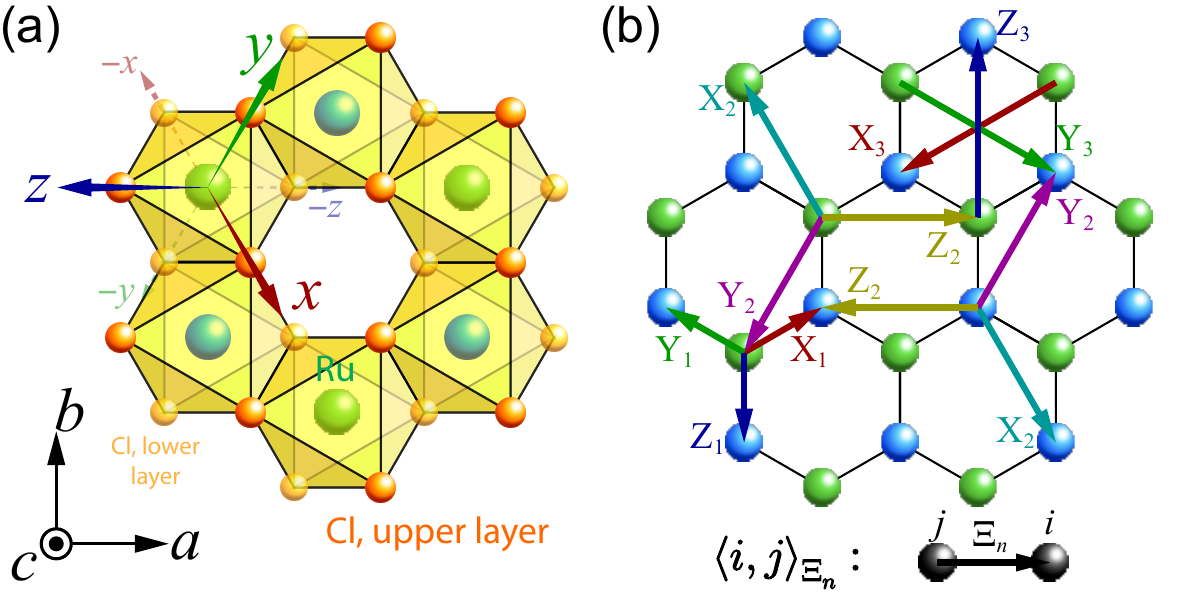}
	\caption{(a) Simplified schematic of the lattice structure of monolayer \ce{\alpha-RuCl_3}, which is a honeycomb lattice of edge-sharing \ce{RuCl_6} octahedrons. Both the chemical frame $xyz$ and the crystallographic frame $abc$ are indicated in the figure. The Cl triangular lattices are actually slightly distorted. (b) Labels of bond types in the hopping term of the Hamiltonian. The blue (green) spheres represent the $A$ ($B$) sublattice of the Ru honeycomb lattice. The notation for the hopping parameters $\Mean{i,j}_{\Xi_n}$ 
 relates panel (b) to terms retained in the hopping Hamiltonian.}
	\label{fig_RuCl3axes_bonds}
\end{figure}

In \ce{\alpha-RuCl_3}, each \ce{Ru^{3+}} ion is at the center of an octahedron formed by 6 \ce{Cl^-} ions, as shown in Fig. \ref{fig_RuCl3axes_bonds} (a). The octahedral crystal field splits the 5 $4d$ orbits of Ru atom into 3 $t_{2g}$ orbits and 2 $e_g$ orbits, with the latter's energy higher by several eV, allowing us to confine our attention to the 
degrees of freedom within the $t_{2g}$ space.
The $t_{2g}$ orbits are stretched along the coordinate planes of the $xyz$ coordinate system shown in Fig. \ref{fig_RuCl3axes_bonds} (a), which differs from the crystallographic $abc$ frame by an orthogonal transformation of basis $\begin{pmatrix} \hat a & \hat b & \hat c \end{pmatrix} = \begin{pmatrix} \hat x & \hat y & \hat z \end{pmatrix} \bfQ$, where
\begin{equation}
	\bfQ = \frac{1}{\sqrt{6}} \begin{pmatrix} 1 & -\sqrt{3} & \sqrt{2} \\ 1 & \sqrt{3} & \sqrt{2} \\ -2 & 0 & \sqrt{2} \end{pmatrix}.
	\label{eq_basis_transform}
\end{equation}
In the insulating case, 5 electrons are present in
the 6-dimensional $t_{2g}$ space on each site so it is more convenient to use a hole representation. 
We adopt the model Hamiltonian from Ref. \cite{winter2016challenges}:
\begin{equation}
	H = H_{\rm int} + H_{\rm SOC} + H_{\rm TB}.
\end{equation}
The interaction part $H_{\rm int}$ is purely on-site and of Kanamori-type \cite{georges2013strong}:
\begin{equation} \begin{array}{c}
	H_{\rm int} = \frac{1}{2} \sum_{i, \xi\eta, \sigma\sigma'} \left( U' h_{i,\xi\sigma}^\dag h_{i,\eta\sigma'}^\dag h_{i,\eta\sigma'} h_{i,\xi\sigma}
	\right. \\ \left. +
	J_H h_{i,\xi\sigma}^\dag h_{i,\xi\sigma'}^\dag h_{i,\eta\sigma'} h_{i,\eta\sigma} +
	J_H h_{i,\xi\sigma}^\dag h_{i,\eta\sigma'}^\dag h_{i,\xi\sigma'} h_{i,\eta\sigma} \right),
\end{array} \end{equation}
where $i$ labels the honeycomb sites, $\xi,\eta = yz(x),xz(y),xy(z)$ label the atomic $t_{2g}$ orbits 
and $\sigma,\sigma' = \up,\down$ label spin. $h_{i, \xi\sigma}^\dag$ ($h_{i, \xi\sigma}$) creates (annihilates) a hole with spin $\sigma$ on the orbit $\xi$ of site $i$. Note that this expression is equivalent to the one given in Ref. \cite{winter2016challenges} with $U' = U - 2J_H$, where $U$ represents the Coulomb interaction strength and $J_H$ is the Hund's coupling amplitude.

In the spin-orbital coupling (SOC) term
\begin{equation}
	H_{\rm SOC} = \lambda \sum_{i,\xi\eta} \left( \LL_{{\rm eff}, \xi\eta} \cdot \SS_{\sigma\sigma'} \right) h_{i,\xi\sigma}^\dag h_{i,\eta\sigma'},
\end{equation}
the elements of the effective orbital angular momentum operator are $L_{{\rm eff}, \xi\eta}^\mu = -i\epsilon_{\mu\xi\eta}$, $\mu,\xi,\eta = x,y,z$ and $\epsilon_{\mu\xi\eta}$ is the Levi-Civita symbol. 
The spin vector operator $\SS$ is half of the Pauli matrix vector.

The spin-independent tight-binding (TB) part
\begin{equation}
	H_{\rm TB} = -\sum_{ij, \xi\eta, \sigma} T_{ij}^{\xi\eta} h_{i,\xi\sigma}^\dag h_{j, \eta\sigma}
\end{equation}
consists of a crystal field (CF) contribution $H_{\rm CF}$ ($i = j$) 
and a hopping contribution $H_{\rm hop}$ ($i\ne j$). 
$H_{\rm CF}$ exists due to the deviation of the true crystal field from a perfect octahedral field, while $H_{\rm hop}$ includes up to the third-nearest neighbor hopping. For bond $\Mean{i,j}_{\Xi_n}$, $T_{ij}^{\xi\eta} = T_{\Xi_n}^{\xi\eta}$ where $\Xi = X, Y, Z$, $n = 1, 2, 3$ and the definitions of all bond types $\Xi_n$ are illustrated in Fig. \ref{fig_RuCl3axes_bonds} (b). In the presence of time-reversal ($\mcT$), spatial inversion ($\mcI$), in-plane 3-fold rotations ($C_3$) 
and a 2-fold rotation symmetry about the $b$ axis ($C_{2b}$), the hopping matrices are constrained to the forms
\begin{equation} \begin{array}{c}
	\bfT_{Z_1} = \begin{pmatrix} t_1 & t_2 & t_4 \\ t_2 & t_1 & t_4 \\ t_4 & t_4 & t_3 \end{pmatrix}, \quad
	\bfT_{Z_2} = \begin{pmatrix} t_1' & t_2' & t_{4'}' \\ t_{2'}' & t_1' & t_4' \\ t_4' & t_{4'}' & t_3' \end{pmatrix}, \\
	\bfT_{Z_3} = \begin{pmatrix} t_1'' & t_2'' & t_4'' \\ t_2'' & t_1'' & t_4'' \\ t_4'' & t_4'' & t_3'' \end{pmatrix},
\end{array} \end{equation}
with all entries real, and $\bfT_{X_n}$ and $\bfT_{Y_n}$ are obtained by successively applying the coordinate rotation $x\rightarrow y\rightarrow z\rightarrow x$ to $\bfT_{Z_n}$. The crystal field matrix is given by
\begin{equation}
	\bfT_{ii} = \begin{pmatrix} 0 & \Delta & \Delta \\ \Delta & 0 & \Delta \\ \Delta & \Delta & 0 \end{pmatrix},
\end{equation}
up to an irrelevant constant.

The parameters of this model have been extracted from bulk \ce{RuCl_3} \textit{ab initio} 
results in Refs. \onlinecite{winter2016challenges, wang2017theoretical}. 
We follow Ref. \onlinecite{winter2016challenges} 
and use $U' = 1.8\rm eV$, $J_H = 0.6\rm eV$ ($U = 3\rm eV$) and $\lambda = 0.15\rm eV$. 
In bulk systems, $C_3$ symmetry is violated by the $C2/m$ layer stacking arrangement.
For the monolayer systems we study below, we recover $C_3$ symmetry 
by averaging the CF Hamiltonian over 3 directions.
For nearest-neighbor hopping processes, we use the $Z_1$-bond parameters previously extracted from \textit{ab initio} simulations of a suspended monolayer \cite{biswas2019electronic}; for further neighbor hoppings we use $Z_2$- and $Z_3$-bond parameters from bulk systems.

\section{Results}
\label{sec_results}

\begin{figure*}[tb]
	\centering
	\includegraphics[width = \textwidth]{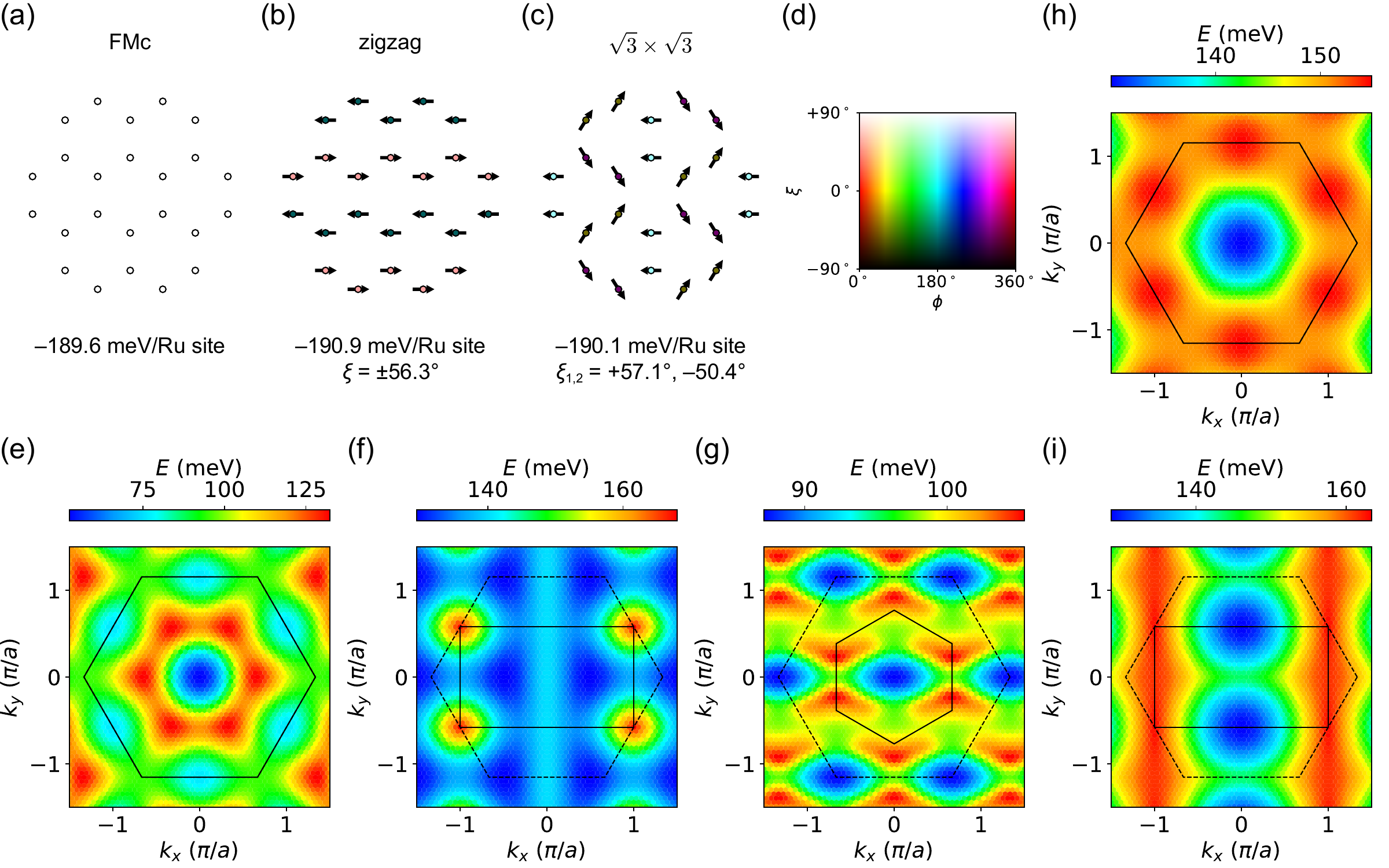}
	\caption{(a)-(c) The atomic magnetic moment configurations of (a) FMc, (b) zigzag and (c) $\sqrt{3} \times \sqrt{3}$ states. The arrows show the in-plane projections of the moments and the colors of the center spots represent the orientation of the moments, with the correspondence illustrated in legend (d), where $\phi$ is the azimuthal angle and $\xi$ is the canting angle, \textit{i.e.} the elevation angle from the plane. Numerical values for energies and canting angles are 
 provided below the figures. (For the $\sqrt{3} \times \sqrt{3}$ state the first canting angle is 
 for the brighter sites and the second for the darker sites.) (e)-(g) The topography of the first conduction band of (e) FMc, (f) zigzag and (g) $\sqrt{3} \times \sqrt{3}$ states. (h)-(i) The topography of the first electronic conduction band of (h) FMc and (i) zigzag states with second- and third-nearest neighbor hopping amplitudes set to zero. In (f), (g) and (i), the dashed hexagon is the structural honeycomb 
 lattice Brillouin zone (BZ) and the solid polygon is the magnetic BZ.}
	\label{fig_bandtopographies}
\end{figure*}

We perform self-consistent Hartree-Fock calculations on the monolayer model, limiting translational symmetry breaking by allowing 
magnetic unit cells with up to $2\sqrt{3} \times 2\sqrt{3}$ honeycomb cells ($12 \times 2 = 24$ Ru sites).
Given the supercell size we choose a variety of initial guesses for the density matrix that allow all other 
symmetries to be broken. 
We first discuss results obtained with the model parameter choices explained in the previous section.
The low-energy magnetic configurations are shown in Figs. \ref{fig_bandtopographies} (a)-(c) and the corresponding lowest
\textit{electronic} conduction bands are shown in Figs. \ref{fig_bandtopographies} (e)-(g).
Of all the extremal states we find, the zigzag state is the lowest in energy.
The out-of-plane ferromagnetic (FMc) state is also competitive but only third lowest,
exceeding the zigzag state by about $1.3\rm meV$ per Ru atom. 
Between them in energy is a $\sqrt{3} \times \sqrt{3}$ state which has not been reported in previous work
and has an energy that is higher than the zigzag state by $0.84\rm meV$ per Ru atom. 
We do not find incommensurate spiral states, despite the strong dominance of $t_2$ which gives rise to a dominant Kitaev term in the spin model under the framework of second-order perturbation theory \cite{winter2016challenges}.
The findings that the ground state is a zigzag and that the FMc state is quite close in energy 
is consistent with experiment, and suggests that the model provides a reasonable description of \ce{RuCl_3}.

To test how robust our results are against changes in model parameters, we also perform the self-consistent Hartree-Fock calculations excluding the second- and third-neighbor hopping integrals. 
As Figs. \ref{fig_bandtopographies} (h)-(i) show, both the conduction bandwidth 
and the momentum-space location of the band bottom are strongly influenced by further-neighbor hopping. 
We will see in Sec. \ref{sec_int_with_graphene} that the location of the conduction band minimum
can have a profound influence on how the band interacts with a proximate layer of graphene. 
In the absence of further-neighbor hopping, the ground state remains zigzag, 
but the energy of FMc state is only $0.11\rm meV$ per Ru site higher, and there is no longer a
competitive $\sqrt{3} \times \sqrt{3}$ solution.  The $\sqrt{3} \times \sqrt{3}$ state is absent 
even if we restore the further-neighbor hoppings to 80\% of their original values, indicating that the state is fragile.

In summary we see that further-neighbor hopping helps to stabilize the zigzag state,
and that it plays an important role in the band structure.  We see in the following section that these relatively 
minor changes in electronic structure have a qualitative influence on how a RuCl$_3$ layer interacts with a 
graphene layer.

\section{Hybridization with Graphene}
\label{sec_int_with_graphene}

The influence of an adjacent insulating van der Waals ferromagnet
layer on the electronic structure of graphene is normally modeled by adding an exchange term to the 
graphene Hamiltonian, which spin-splits its bands. If the adjacent layer is AFM, as in the RuCl$_3$ case,
the influence of oppositely directed moments tends to cancel because electrons in a small graphene Fermi pocket tend to 
average over atomic length scales.  The situation can be more complex, however, when there is charge transfer 
between graphene and the van der Waals magnet, something that is known to occur \cite{rizzo2020chargetransfer, wang2020modulation}
in the RuCl$_3$ case.  
In discussing the electronic structure of the graphene/\ce{RuCl_3} system we recognize that although the 2D materials share a triangular Bravais lattice, they have very different Bravais lattice
constants ($a_{\rm G} = 2.46\text{\AA}$ for the graphene layer and $a_{\rm R} = 5.8\text{\AA}$ for the 
RuCl$_3$) and their relative orientations are normally uncontrolled.  In the rest of this section, we first model the interlayer 
coupling and then analyze its influence on the electronic structure of the combined system.

We assume a spin-conserving tunneling amplitude $T_\xi(\rr)$ from $t_{2g}$ $d$-orbital 
$\xi$ in \ce{RuCl_3} to the $p_c$ orbital
\footnote{
We refer to the low-energy bands of graphene as $p_c$ bands
because we have denoted the crystallographic frame by $abc$ rather than $xyz$.} in graphene that is some function of the difference $\rr$ 
between the lateral 2D positions of the atoms in question.
At this point we do not specify the concrete form of the function $T_\xi(\rr)$. Later on we will see that a model 
can be constructed from a small number of phenomenological parameters related to these functions.

\begin{figure*}[tb]
	\centering
	\includegraphics[width = \textwidth]{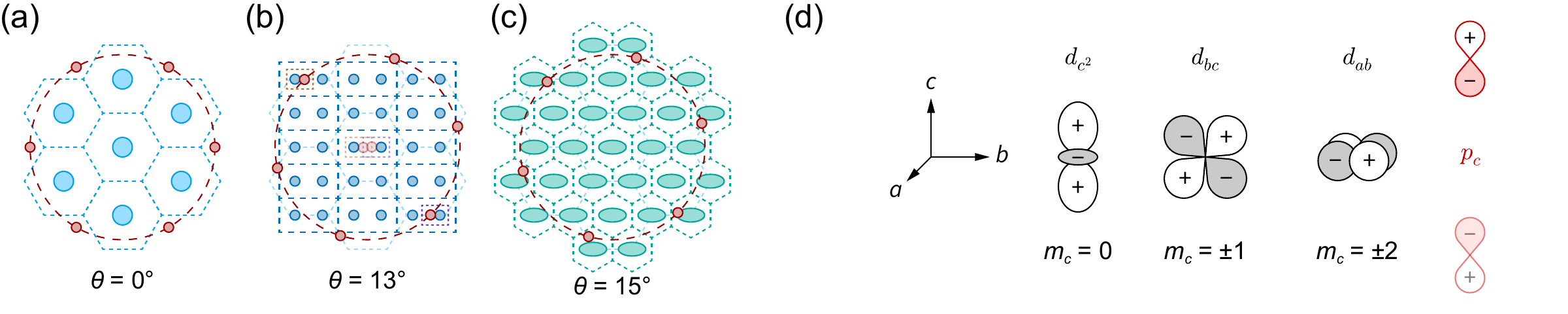}
	\caption{(a)-(c) Schematics of isolated graphene (red) and \ce{RuCl_3} Fermi surfaces (light blue, blue or green) and their reciprocal-lattice replicas for (a) FMc, (b) zigzag and (c) $\sqrt{3} \times \sqrt{3}$ magnetic states.
These figures roughly assume that the electron charge transfer from graphene to \ce{RuCl_3} is $0.04e$ per Ru atom 
($0.007e$ per carbon atom, $2.7 \times 10^{13} \rm cm^{-2}$ \cite{rizzo2020chargetransfer}). 
The illustrated twist angle of the graphene lattice with respect to \ce{RuCl_3} is indicated below each figure. 
The bands of the FMc and $\sqrt{3} \times \sqrt{3}$ states are non-degenerate whereas those of the 
zigzag state are doubly degenerate.  It follows that for the FMc and $\sqrt{3} \times \sqrt{3}$ states the total Fermi surface area
per magnetic BZ is double that of graphene, whereas that of the zigzag state is the same.  
However, the fraction of $\bm{k}$-space that is within
a replica Fermi surface is equal in the FMc and zigzag cases, because the zigzag BZ is half as large,
while the replicated $\sqrt{3} \times \sqrt{3}$ Fermi surfaces occupy an area that is three times larger because
the BZ area of this state is $1/3$ of that of FMc.
In (b), the intersecting Fermi surfaces are highlighted by small dashed boxes, 
and their replicas in the first BZ of \ce{RuCl_3} are also shown.
(d) Schematics of some $d$ orbits in the $abc$ frame of \ce{RuCl_3} and the $p_c$ orbit of graphene.}
	\label{fig_FS}
\end{figure*}

Following a procedure similar to that used in the derivation of Bistritzer-MacDonald model of twisted bilayer graphene \cite{bistritzer2011moire}, we obtain the following expression for interlayer tunneling between Bloch states in the two layers:
\begin{equation} \begin{array}{c}
\label{eq:tunnelingme}
	\Braoperket{{\rm G}, \alpha\sigma\kk}{H}{{\rm R}, \beta\xi\sigma'\pp} = \delta_{\sigma\sigma'} \times \\
	\sum_{\GG^{\rm G}, \GG^{\rm R}} 
 w_\xi(\kk + \GG^{\rm G}) \, e^{i \left( \GG^{\rm G} \cdot \ttau_\alpha^{\rm G} - \GG^{\rm R} \cdot \ttau_\beta^{\rm R} \right)} \, \delta_{\kk + \GG^{\rm G}, \pp + \GG^{\rm R}},
\end{array} \end{equation}
where $\alpha$ ($\beta$) labels the graphene (\ce{RuCl_3}) sublattice, 
$\sigma,\sigma' = \up,\down$ label spin, $\GG^{\rm G}$ and $\GG^{\rm R}$ are respectively reciprocal lattice vectors of the graphene and \ce{RuCl_3} layer, $\ttau_\alpha^{\rm G}$ ($\ttau_\beta^{\rm R}$) is the location of 
sublattice $\alpha$ ($\beta$) in a graphene (\ce{RuCl_3}) unit cell, and
\begin{equation}
	w_\xi(\qq) = \frac{1}{\sqrt{\Omega_c^{\rm G} \Omega_c^{\rm R}}} \int d^2\rr T_\xi(\rr) e^{-i\qq\cdot\rr}
\end{equation}
is proportional to the Fourier transform of the real-space tunneling function.
$\Omega_c^{\rm G}$ and $\Omega_c^{\rm R}$ are the unit cell areas of graphene and \ce{RuCl_3}, respectively. 
We see that the condition for Bloch states in the two layers to hybridize
is that the two momenta should be equal when reduced to the BZ of one layer
up to a reciprocal lattice vector of the other layer.

Because the separation between layers exceeds the atom size in either layer,
$w_\xi(\qq)$ typically drops rapidly with $|\qq|$.
We are mainly interested in how hybridization with the magnetic insulators influences the transport properties of 
graphene, and therefore mainly interested in single-particle states with energies close to
the graphene Fermi energy.  Because the carrier density in graphene layer is relatively small when normalized per atom,  
its low energy electronic states lie at momenta close to the Dirac points 
$\pm\KK^{\rm G} = \pm (4\pi/3a_{\rm G}) (\cos\theta, \sin\theta)$. 
(Here $\theta$ is the orientation angle of graphene relative to that of \ce{RuCl_3}.)
For this reason we can replace $w_\xi(\kk + \GG^{\rm G})$ in Eq.~\ref{eq:tunnelingme} by $w_\xi(\KK)$, 
where $\KK$ is the closest graphene BZ corner.  {\it Ab initio} electronic structure calculations \cite{biswas2019electronic} suggest a typical hybridization strength $w_\xi(\KK) \sim 10\rm meV$.  The RuCl$_3$ BZ momenta ${\bm{p}}$ that satisfy 
the $\delta$-function in Eq. (\ref{eq:tunnelingme}) are those that are close to a corner  
when reduced to graphene's BZ.  We discuss the implications of these properties further below.  

We discuss two limits of coupling between graphene's near-Fermi surface states and \ce{RuCl_3} states.  When the graphene state couples to a \ce{RuCl_3} state that is separated from the Fermi energy by an energy $\Delta E$
that is much greater than the hybridization scale $w_\xi(\KK)$, the hybridization can be treated perturbatively and 
gives rise to energy shifts that are $\sim w_{\xi}(\KK)^2/\Delta E$ and potentially spin-dependent as 
discussed further below.  These spin-dependent energy shifts are the exchange interactions between
graphene quasiparticles and the magnetic insulator spins mentioned above.
On the other hand if a \ce{RuCl_3} state is within $w_\xi(\KK)$ around the Fermi energy, 
it will couple strongly to the graphene orbitals, creating avoided crossing gaps and reducing
graphene orbital velocities to cause a significant drop in the conductivity of the system. 
Measurements of graphene transport properties 
therefore can be used to determine when the graphene Fermi surface intersects
or nearly intersects the \ce{RuCl_3} Fermi surface. 

The carriers of the Mott insulator \ce{RuCl_3} can be of either magnetic polarons or Fermi liquid 
character type depending on the level of doping \cite{lee2006doping, keimer2015quantum, koepsell2021microscopic}. 
Here we assume that the location and shape of Fermi surface \ce{RuCl_3} is captured by  
rigid-band occupation of the conduction band obtained in Hartree-Fock calculations
for the insulating state.  
As mentioned previously, those mean-field theory calculations do 
describe the competition between magnetic configurations reasonably accurately.
As we explain below, the coupling between \ce{RuCl_3} and graphene layers can be strongly
sensitive to their relative orientation $\theta$. If $\theta$ is not controlled when the 
bilayer device is fabricated, this sensitivity will lead to strongly device-dependent properties.
On the other hand, {\it in situ} twist angle control \cite{ribeiro-palau2018twistable, inbar2023quantum} could add to
the power of graphene transport probes of insulating magnetic states.
In our analysis we allow for arbitrary relative orientation between the layers.

As Figs. \ref{fig_FS} (a)-(c) show, Fermi surface intersections do not occur
for any relative twist angle when \ce{RuCl_3} is in the FMc state. 
In contrast, for the zigzag state, we estimate that they occur within the twist angle range $\theta \sim (17 \pm 4)^\circ$.
For the $\sqrt{3} \times \sqrt{3}$ state, intersections occur
within both $(0 \pm 5)^\circ$ and $(16 \pm 6)^\circ$ orientation intervals.
In general the larger the magnetic unit cell, the smaller the magnetic BZ, 
the denser the magnetic insulator Fermi surface $\kk$ space replicas, increasing 
the chance that Fermi surfaces intersect.  When the insulator has an ordered magnetic state, lower 
translational symmetry (larger magnetic unit cells) limits the ability of momentum conservation 
to restrict hybridization. We speculate 
that interlayer hybridization has a weaker effect in spin-liquid states because they do not 
break translational symmetry.  The role of interlayer hybridization in graphene/Kitaev spin liquid heterostructures
could be explored by generalizing earlier Kondo-Kitaev lattice model analyses
\cite{leeb2021anomalous, jin2021flat} to the physically relevant 
case of incommensurate lattices, but this is outside the scope of the current study.


\begin{figure*}[tb]
	\centering
	\includegraphics[width = \textwidth]{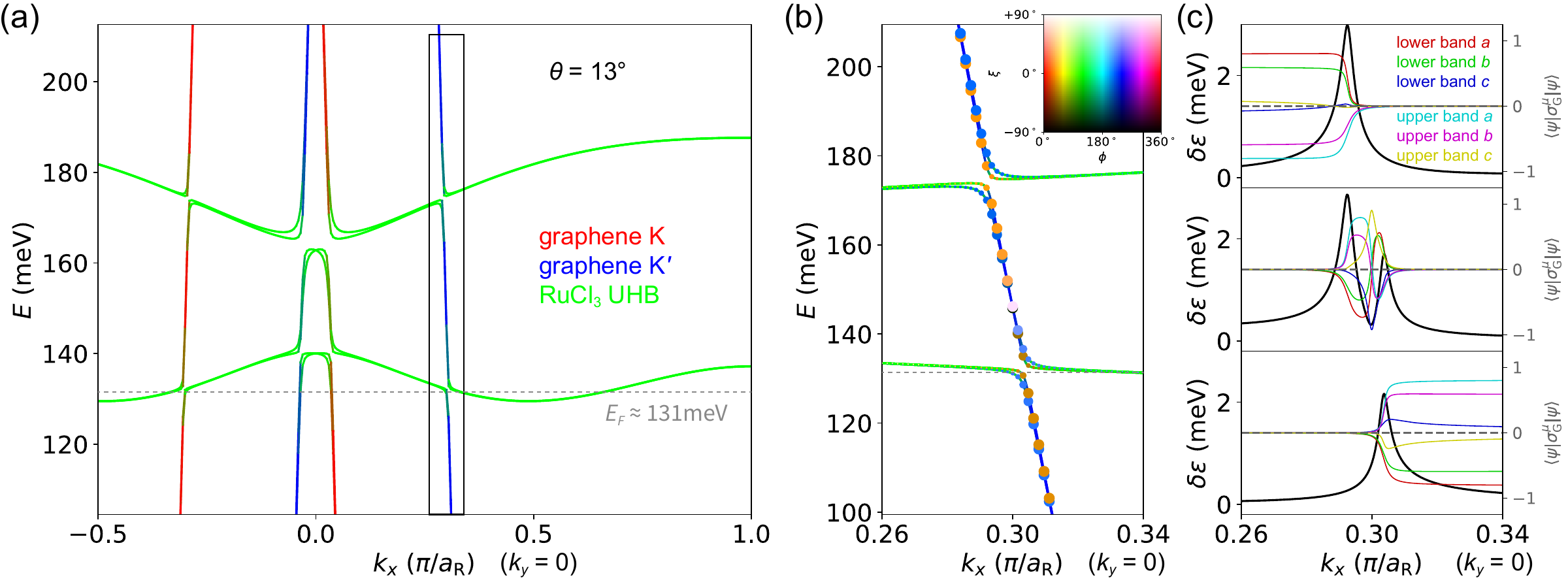}
	\caption{(a) Close up of an avoided crossing between 
strongly dispersing graphene and weakly dispersing \ce{RuCl_3} bands for the zigzag state of 
\ce{RuCl_3}.  As is clear from Fig. \ref{fig_FS} (b), at orientation $\theta = 13^{\circ}$ 
the Fermi surfaces of the two graphene valleys have a small intersection
near $\bm{k}=0$ when mapped to the zigzag state magnetic BZ.  
For this illustration, we have taken the energy of graphene's Dirac point  
to be $650\rm meV$ higher than the upper Hubbard band bottom, $w_0 = 10\rm meV$, $w_1 = 5\rm meV$ and $w_2 = 3\rm meV$. 
We see that near \ce{RuCl_3}'s BZ center, its states couple to both valleys of graphene. 
(b) An enlarged illustration of the avoided band crossing from the boxed region in (e).
The dashed lines in (a) and (b) mark the position of the Fermi energy. 
The size of the spots in (b) represents the graphene fraction of the eigenstates, and the color represents the spin orientation of the graphene component of the eigenstate, using the color scale shown at the top right corner (and also in Fig. \ref{fig_bandtopographies} (d)).  The graphene spin splitting is largest close to the avoided crossing points, and the spin orientations of the spin-split states lie close to the $ab$ plane.  
(c) Plot of Kramers degeneracy splitting $\delta \epsilon$ and the expectation values of 
the graphene-projected spin Pauli matrices $\sigma_{\rm G}^\mu$ in the three spatial directions $\mu = a, b, c$. 
The upper, middle and lower panels correspond respectively to the upper two, middle two and lower two bands shown in (b). We see that close to the hybridization points, the spin splitting can be up to $\sim 3\rm meV$, while away from the hybridization points the exchange splitting of graphene energies rapidly goes down to $\sim 0.1\rm meV$ with the splitting of zigzag upper Hubbard bands even weaker. The spin-split graphene bands generally have almost in-plane spin orientations, which tend to stabilize at
azimuthal angles close to $36^\circ$ 
and $216^\circ$ as one goes away from hybridization point.}
	\label{fig_hybridize}
\end{figure*}

There is an opportunity to reduce the number of free parameters that are relevant for 
hybridization by symmetry. This can be seen by noting that the $t_{2g}$ orbital wave functions, $\phi_{yz}(\xx) = yz\Phi(|\xx|)$, $\phi_{xz}(\xx) = xz\Phi(|\xx|)$ and $\phi_{xy}(\xx) = xy\Phi(|\xx|)$, are 
linear combinations of the $c$-projected angular momentum $m_c$ eigenstates:
\begin{equation} \begin{cases}
	\phi_{m_c=0}(\xx) = \frac{1}{\sqrt{3}} \left( c^2 - \frac{a^2+b^2}{2} \right) \Phi(|\xx|), \\\addlinespace
	\phi_{m_c=\pm1}(\xx) = \frac{(a\pm i b)c}{\sqrt{2}} \Phi(|\xx|), \\\addlinespace
	\phi_{m_c=\pm2}(\xx) = \frac{(a\pm i b)^2}{2\sqrt{2}} \Phi(|\xx|).
\end{cases} \end{equation}
Here we use $\xx$ to represent 3D position vectors to distinguish from 2D position vector $\rr$.  
In a two-center approximation,
$T_{m_c}(\rr) = T_{\Abs{m_c}}(\Abs{\rr}) e^{im_c\theta_{\rr}}$, implying that
\begin{equation}
	w_{m_c} \left( \qq \right) = i^{m_c} w_{\Abs{m_c}}(\Abs{\qq}) e^{i m_c \theta_{\qq}}.
\end{equation}
In this way we are able to express interlayer coupling in terms of three real
parameters $w_{m_c} = w_{m_c}(\Abs{\KK^{\rm G}})$, where $m_c = 0, 1, 2$. 
We expect the energy hierarchy between them to be $|w_0| > |w_1| > |w_2|$ since from the schematic shown in Fig. \ref{fig_FS} (d), larger $\Abs{m_c}$ orbitals orbit extend more along $ab$ plane 
and less along $c$ axis, implying that 
the real-space tunneling function $T_{m_c}(\rr)$ has broader range but overall smaller absolute value, and thus that 
the momentum-space tunneling function $w_{m_c}(\qq)$ has narrower range. In Fig.~\ref{fig_hybridize} (a) we take $w_0 = 10\rm meV$, $w_1 = 5\rm meV$ and $w_2 = 3\rm meV$ and graphene Dirac velocity $\hbar v_F = 7\rm eV \cdot \text{\AA}$.
These numbers have only a qualitative justification, especially when the possible importance of hopping between 
metal ions via \ce{Cl} p-bands is recognized.  Their order of magnitude is chosen to reproduce the typical size of \textit{ab initio} band hybridization gaps.

In the decoupled limit, both the graphene and the zigzag state \ce{RuCl_3} bands are doubly degenerate.
Both degeneracies can be viewed as Kramer's degeneracies protected by 
$\kk$-preserving $\mcI\mcT$ symmetry, where $\mcI$ is the 3D spatial inversion operator and $\mcT$ is the time-reversal operator.
When the coupling between the two layers is turned on, the band degeneracies of both materials are weakly lifted.
This observation is consistent with the fact that placing graphene on top of a \ce{RuCl_3} layer breaks overall
$\mcI\mcT$ symmetry. Because its tunneling amplitudes are odd under reflection through a plane 
midway between the two layers, the $w_1$ contribution dominates the Kramers violation.  
Away from the hybridization points, the exchange splitting of the graphene bands quickly drops to values 
$\sim 0.1\rm meV$, far smaller than in a graphene/FM insulator heterostructure.
This finding is consistent with our expectation that the exchange coupling effects in graphene are 
extremely small when the magnetic layer does not have a net moment.  
At the hybridization points, both spin channels of the graphene band are gapped
and the spin splitting increases to $2\sim 3\rm meV$.
We therefore expect that a reduction in overall electrical conductivity will accompany 
Fermi surface intersections, even when the magnetic insulator is in an AFM state.
Note that in the coupled bilayer system the total spin-magnetization of the combined system in nominally
AFM systems will generically be non-zero because of the small splitting of polarized states at the 
Fermi energy.  

The case in which the magnetic layer has aligned spins, either in its ground state as in 
the FM insulator \ce{CrI_3} \cite{tseng2022gatetunable} or due to field alignment, is distinct.
The bands of the isolated magnetic layer are then non-degenerate and spin-polarized, so that 
only one graphene spin-component is influenced by hybridization.  
For example in the case of graphene/\ce{CrI_3}, the spin-degenerate graphene $\pi$-bands 
cross \cite{cardoso2018van, cardosostrong} a non-degenerate band of the magnetic insulator.  When hybridization is included,
only one graphene spin channel is gapped.  Under these circumstances we 
expect spin-polarized electronic transport, with the conductivity of the graphene 
spin-component that is present in the conduction band of the magnetic layer suppressed.
We predict, based on our electronic structure model, that this suppression is 
anomalously weak in spin-aligned \ce{RuCl_3}, independent of twist angle, because of the 
absence of Fermi surface intersections illustrated in Fig.~\ref{fig_FS}(a).
In contrast, graphene spin-transport would be expected to be strongly spin-polarized 
in graphene/\ce{RuCl_3} if the fragile $\sqrt{3} \times \sqrt{3}$ magnetic state were stable, since 
this configuration has Fermi surface intersections at many twist angles.

We note that in some spin-aligned materials, including in FM \ce{CrI_3},
the conduction band quasiparticles have the majority spin, differing from valence band states via orbital 
instead of spin quantum numbers.
This is unlike the \ce{RuCl_3} case, in which the conduction band states can be viewed as 
forming an upper Hubbard band with spins opposite to those of the occupied valence band states. 
In the hypothetical case in which the combined system Kramer's degeneracy is protected by $C_2\mcT$ symmetry,
where $C_2$ is the in-plane two-fold rotation, interlayer coupling would not break Kramer's degeneracy and 
transport would not be spin-dependent, but conductivity suppression will still accompany
Fermi surface intersections.

\section{Summary and Discussion}
\label{sec_summary}

In this work we have analyzed how the electronic properties of an adjacent graphene layer can be used to 
probe the magnetic state of 2D magnetic insulators, focusing on the Kitaev material
\ce{RuCl_3} as a typical interesting example.
We assume that the Dirac point of graphene lies outside the band gap of 
the insulator so that charge transfer occurs between the two single-layer 2D systems.
Charge transfer occurs in Cr, Fe and Ru trihalides \cite{zhang2018strong, cardoso2018van, rizzo2020chargetransfer, wang2020modulation, li2020geometric, tseng2022gatetunable, lyu2022giant, cardosostrong}, and we believe that it is likely to 
be common.  We predict that hybridization of the materials can lead to sizable magneto-resistance 
that is sensitive to the magnetic configuration of the insulator and to the relative orientation of the two layers.

Our specific predictions rely on a mean-field model description of both graphene and 
the doped 2D magnetic insulator.  We find that Hartree-Fock theory applied to a simple but realisitic  
model Hamiltonian is able to predict the relative stability of 
different magnetic configurations of the single-layer Mott insulator \ce{RuCl_3}. 
In agreement with earlier work \cite{sarikurt2018electronic, gerber2020initio, souza2022magnetic},
we find that the ground state is a zigzag AFM but the out-of-plane FM state is also competitive --
explaining the weak fields required to achieve spin alignment in bulk \ce{RuCl_3}.  
We also find a state with a $\sqrt{3} \times \sqrt{3}$ magnetic cell that is metastable in a narrow range of model parameters but is very fragile.  

Our explicit calculations have a number of deficiencies.  First of all, we do not account for many-body fluctuations in 
Bloch state occupation numbers.  These fluctuations can in principle \cite{lee2006doping} eliminate the momentum-space occupation
number discontinuities associated with the Fermi surfaces we have discussed.  We do not, however, expect them to 
change the locations in momentum space that have high spectral weight near the Fermi energy.  It follows that 
the conclusions of our mean-field analysis should, for the most part, be unaffected. 

Secondly, we do not account for 
structural relaxation of the van der Waals bilayers.  Atomic position readjustments
will be larger in the RuCl$_3$ layer, which is not as stiff as the graphene layer, but 
we do not expect them to be large because interlayer interactions are weak.
The \ce{RuCl_6} crystal field in isolated monolayer \ce{RuCl_3} is close to the ideal octahedral form, as can be seen from the relatively small values of $\Delta$ and $t_4$ in the model \cite{biswas2019electronic}.  We expect the relaxations of the Cl ions closer to the graphene layer to distort the metal coordination further from the octahedral ideal than in the isolated layer case, which could  
increase the contribution of Ru--Cl--Ru indirect hopping to the nearest-neighbor hopping parameters \cite{winter2016challenges}. 
Reference \onlinecite{gerber2020initio} predicts that the effect of graphene favors the zigzag state, 
and also makes other AFM states more competitive.

We analyze the coupled bilayer within a mean-field framework by considering the effect of interlayer hybridization as a 
weak perturbation.  For \ce{RuCl_3} the hybridization energy scale is estimated to be $w \sim 10$ meV, smaller than the 
bandwidth of either material.
We argue that its influence on electronic properties is sensitive primarily to the presence or absence of 
intersections between the Fermi surfaces of the two materials when the magnetic material Fermi surface is 
plotted in an extended-zone scheme as in Fig.~\ref{fig_FS}.
Intersections between Fermi surfaces will always suppress the graphene conductivity, and 
in the case of spin-aligned magnetic insulators, make it spin-dependent.  
Intersections occur for both zigzag and $\sqrt{3} \times \sqrt{3}$ states of \ce{RuCl_3} over certain ranges of relative orientation angle between the two layers.  In the zigzag case both spin channels of graphene are gapped.
Spin-splittings $\delta \epsilon \lesssim w$ occur only at those points in momentum space 
where strong hybridization occurs.  Upon moving away from band crossing points, 
the symmetries of the isolated layers are gradually recovered and the spin-splitting is very small.
On the other hand for out-of-plane spin-aligned magnetic configurations, weak exchange splitting occurs throughout momentum space.
In the case of \ce{RuCl_3}, these weakly momentum dependent exchange splittings are the main consequence of 
hybridization since Fermi surface intersections do not occur. 

We have discussed mainly the influence of coupling between layers on lateral transport within the 
graphene layers, but our theory can also be used to address vertical transport.
However, it may not be clear exactly how this quantity is best measured because it may be difficult to contact the 
magnetic insulator layers. The more interesting observable is, perhaps, transport between graphene layers through a 
magnetic layer. For both vertical and lateral transports, 
fabrication of graphene/magnetic-insulator devices with {\it in situ} twist-angle-control \cite{ribeiro-palau2018twistable,inbar2023quantum} 
would greatly enhance the power of transport probes,
because it would allow the Fermi surface of the magnetic insulator to be mapped out.

Although we have focused on the influence of the magnetic state on graphene transport, which is more readily observable, 
there must also be inverse effect in which hybridization 
tilts the competition between magnetic states. On general grounds interlayer hybridization will lower energy and 
therefore favor ordered states that have large Fermi surface overlaps with graphene.  This effect will be small
however and can only be the deciding factor if the pre-hybridization energies of the competing states are very close.
It should also favor ordered magnetic states over spin-liquid states, in which inter-layer hybridization 
is likely suppressed by quasiparticle fractionalization.
Like all hybridization effects, the influence on 
magnetic state competitions can be enhanced by applying pressure to narrow the van der Waals layer separations.
This strategy might be readily simple to purse experimentally.  
Interesting future directions also include explorations on how transport properties of graphene can probe the magnons of ordered states \cite{yang2023magnetic} and the magnetic polarons of the doped Mott-insulating states of substrate.

\begin{acknowledgments}
This work was supported by the U.S. Department of Energy, Office of Science, Basic Energy Sciences, under Award DE-SC0022106.
We gratefully acknowledge helpful discussions with James Analytis, Kwabena Bediako, Ken Burch, Johannes Knolle, Eslam Khalaf, Ziyu Liu, Ipsita Mandal, Jonathon Nessralla, Joaquin Fernandez-Rossier, Dihao Sun, and Roser Valent\'i.  This work was enabled by computational resources 
provided by the Texas Advanced Computing Center. 
\end{acknowledgments}


\bibliographystyle{apsrev4-1}
\bibliography{bibliography}

\end{document}